\documentclass[sigconf,authorversion,nonacm]{acmart}

\acmConference[FSE 2024]{International Symposium on the Foundations of Software Engineering}{15 -- 19 July 2024}{Porto de Galinhas, Brazil}

\usepackage{amsmath,amsfonts}
\usepackage{subcaption}
\usepackage{algorithmic}
\usepackage{graphicx}
\usepackage{textcomp}
\usepackage{xcolor}
\usepackage{url}
\usepackage{microtype}

\setcopyright{acmlicensed}
\acmYear{2024}
\copyrightyear{2024}
\acmVolume{1}
\acmNumber{FSE}
\acmArticle{48}
\acmMonth{7}

\usepackage{xcolor}
\usepackage{xspace}

\newcommand{\etal}{\hbox{\emph{et al.}}\xspace}

\newcommand{\ie}{\hbox{\emph{i.e.}}\xspace}

\PackageWarning{TODO:}{X}
\PackageWarning{TODO:}{X\%}

\usepackage{tcolorbox}

\usepackage{tcolorbox}

\newlength{\imagewidth}
\usepackage{xspace}
\usepackage{dirtytalk}
\usepackage{xcolor}

\newcommand{\cc}{\textsc{CodeCompose}\xspace}
\newcommand{\company}{Meta\xspace}
\newcommand{\Meta}{Meta\xspace}
\newcommand{\Bento}{Bento\xspace}

\usepackage{xspace}
\usepackage{dirtytalk}
\usepackage{xcolor}

\definecolor{teal}{RGB}{0,128,128}

\newcommand{\DefMacro}[2]{\expandafter\newcommand\csname rmk-#1\endcsname{#2}}
\newcommand{\UseMacro}[1]{\csname rmk-#1\endcsname}
\newcommand{\rom}[1]{\uppercase\expandafter{\romannumeral #1\relax}}

\begin{document}

\author{Omer Dunay} \email{omerdu@meta.com}
\affiliation{%
  \institution{Meta Platforms, Inc.}
  \city{Menlo Park, CA}
  \country{USA}}
  
\author{Daniel Cheng} \email{danielcheng@meta.com}
\orcid{0009-0007-1708-8132}
\affiliation{%
  \institution{Meta Platforms, Inc.}
  \city{Bellevue, WA}
  \country{USA}}

\author{Adam Tait}
\email{adamtait@meta.com}
\affiliation{
  \institution{Meta Platforms Inc.}
  \country{USA}
}
  
\author{Parth Thakkar} \email{parthdt@meta.com}
\affiliation{%
  \institution{Meta Platforms, Inc.}
  \city{Menlo Park, CA}
  \country{USA}}

\author{Peter C. Rigby} \orcid{0000-0003-1137-4297} \authornote{Rigby is a professor at Concordia University in Montreal, QC, Canada.}  \email{pcr@meta.com}
\affiliation{%
  \institution{Meta Platforms, Inc.}
  \country{USA}}
  
\author{Andy Chiu} \email{achiu@meta.com}
\affiliation{%
  \institution{Meta Platforms, Inc.}
  \country{USA}}
  
\author{Imad Ahmad} \email{imadahmad@meta.com}
\orcid{0009-0005-8606-3946}
\affiliation{%
  \institution{Meta Platforms, Inc.}
  \city{Menlo Park, CA}
  \country{USA}}
  
\author{Arun Ganesan} \email{arunganesan@meta.com}
\affiliation{%
  \institution{Meta Platforms, Inc.}
  \country{USA}}
    
\author{Chandra Maddila} \email{cmaddila@meta.com} 
\orcid{0000-0002-9432-1045}
\affiliation{%
  \institution{Meta Platforms, Inc.}
  \city{Bellevue, WA}
  \country{USA}}
    
\author{Vijayaraghavan Murali} 
\orcid{0009-0009-1374-7334}
\email{vijaymurali@meta.com}
\affiliation{%
  \institution{Meta Platforms, Inc.}
  \city{Menlo Park, CA}
  \country{USA}}

\author{Ali Tayyebi} \email{alitayyebi@meta.com}
\affiliation{%
  \institution{Meta Platforms, Inc.}
  \city{New York, NY}
  \country{USA}}

\author{Nachi Nagappan} \email{nnachi@meta.com}
\affiliation{%
  \institution{Meta Platforms, Inc.}
  \country{USA}}

\title{Multi-line AI-assisted Code Authoring}

\begin{abstract}
\cc is an AI-assisted code authoring tool powered by large language models (LLMs) that provides inline suggestions to 10's of thousands of developers at \company. In this paper, we present how we scaled the product from displaying single-line suggestions to multi-line suggestions. This evolution required us to overcome several unique challenges in improving the usability of these suggestions for developers.

First, we discuss how multi-line suggestions can have a "jarring" effect, as the LLM's suggestions constantly move around the developer's existing code, which would otherwise result in decreased productivity and satisfaction. 

Second, multi-line suggestions take significantly longer to generate; hence we present several innovative investments we made to reduce the perceived latency for users. These model-hosting optimizations sped up multi-line suggestion latency by 2.5x.

Finally, we conduct experiments on 10's of thousands of engineers to understand how multi-line suggestions impact the user experience and contrast this with single-line suggestions. Our experiments reveal that (i) multi-line suggestions account for 42\% of total characters accepted (despite only accounting for 16\% for displayed suggestions) (ii) multi-line suggestions almost doubled the percentage of keystrokes saved for users from 9\% to 17\%. Multi-line \cc has been rolled out to all engineers at \Meta, and less than 1\% of engineers have opted out of multi-line suggestions.

\end{abstract}


\maketitle

\section{Introduction}


\cc~\cite{codecompose2024} provides inline suggestions as a software engineer types code, but it was originally only designed to predict tokens that would complete the current line. Such single-line suggestions should be quick, highly accurate, and help with the immediate context. In contrast, multi-line suggestions need to be deeply informed and intelligent. These code blocks can help users discover APIs, best practices, and implementation details. In this work, we describe how we added multi-line suggestions to \cc. 

\cc suggests code as the user types, impacting the main authoring workflow.
Multi-line suggestions can have a “jarring” effect as the displayed suggestions move around code that the user has already written. This mingling of suggested code with human written code has a high cognitive load on the user, as they have to review and rework code that they trust and wrote, with code that they need to verify. We adhered to strict design principles of not being intrusive, and of fitting seamlessly into the existing user experience.

There were three main challenges in developing multi-line suggestions at scale: (1) eliminating the jarring effect (2) providing responsive suggestions with low latency for large blocks of code to be generated (3) rolling out and evaluating the impact of multi-line to 10's of thousands of developers.

\textbf{Challenge 1. User interface experience - Eliminating the ``jarring'' effect}

Avoiding the ``jarring” effect is simple with single-line suggestions:
we only show suggestions when the cursor is at the end of a line (with the exception of special characters at the end like brackets).
In contrast, multi-line is complex as it can disrupt code that is already written. Creating an algorithm for this solution poses a non-trivial technical problem, including figuring out how to determine the context of the cursor reliably. Moreover responses from the LLM can be not well formatted, and aligning them into the cursor position may cause further distraction for the user. 

\cc handles these challenges with a combination of pre-processing and post-processing algorithms that utilize a semantic context understanding of the programming language and cursor position scope. 
Section~\ref{sec:jarring} discusses this multi-line algorithm based on the semantic scope to minimize noise to the user.

\textbf{Challenge 2.  Responsive User Experience}

\cc's inline suggestions show up automatically as the user types. Each suggestion aligns with a given state of the active file, and is being invalidated as soon as the user types an additional keystroke or moves their cursor. Multi-line suggestions are long by nature; therefore it may take a few seconds for the LLM to generate them, during which time the user may hit another keystroke and dismiss the response before they have even seen the suggestion.

Since latency is a key factor in determining the "display rate", \ie the number of suggestions that users are actually able to view, we invested in reducing the latency of long multi-line suggestions through both the client extension and the model-hosting service, as outlined in Section~\ref{sec:model}.
%

\textbf{Challenge 3. Production release and effectiveness measurement}

The final challenge we faced was in monitoring the effectiveness of the experimental features enumerated in the prior sections. We tracked a host of metrics online to evaluate whether each feature was improving the user experience, and adjusted the product accordingly. In particular, we needed to evaluate whether users found multi-line suggestions more useful compared to only single-line suggestions, due to the higher latency and reduced display rate of multi-line suggestions.

These metrics included acceptance rate, display rate, latency, \% keystrokes saved, and \# of characters accepted per user. These were broken down by single-line vs multi-line suggestions to determine the net benefit of each feature for users. 

\textbf{Result Summary.} In monitoring our online A/B tests, we found that the investment in multi-line suggestions disproportionately increased throughput, as:
\begin{itemize}
    \item Multi-line suggestions accounted for 42\% of total characters accepted day (despite only accounting for at 16\% of displayed suggestions)
    \item Multi-line suggestions drove a significant net increase in \% keystrokes saved from 9\% to 17\%, as shown in Table~\ref{fig:keystrokes}
\end{itemize}

This paper is structured as follows. In Section~\ref{sec:background}, we give background on \Meta and introduce our experimental methodology. In Sections~\ref{sec:jarring} to \ref{sec:production}, we address each challenge. In Section~\ref{sec:threats}, we discuss threats to validity. In Sections~\ref{sec:literatureAndDiscussion} and \ref{sec:conclusion}, we discuss related work and conclude the paper. 


\section{Background and Methodology}
\label{sec:background}

\subsection{\company}
\company is a major industrial software organization with a complex code base that covers a wide range of applications ranging from social networking and virtual reality to software engineering infrastructure, such as continuous integration tooling and workplace coordination. Thousands of developers work on billions of lines of code (LOCs) in a monolithic repository that contains source code written in tens of programming languages. At \company, code authoring is one of the prominent activities in the Software Development Life Cycle (SDLC). A significant amount of contextual knowledge on the internal software development processes, and libraries, is contained within in the source code or is confined to a small set of developers. For example, a question like \emph{``How can I query a table from Hive in Hack?''} can be answered using that knowledge.
The state of the source code and the developers associated with the source code are in a constant state of evolution - internal frameworks and tools get added and deprecated, while developers move across teams and change roles.
At \company's scale, keeping up with the knowledge required to accomplish coding tasks is a major challenge. Additionally, the dynamic environment at a large software company like \company poses several challenges concerning knowledge discovery and improving developer productivity. 

\subsection{\cc at \company}
At \company, we built an AI-assisted code authoring system named \cc to explore the application of LLM technology for code authoring~\cite{codecompose2024}.
Our team conducted R\&D on the underlying LLM architecture and converged on using InCoder-6.7B~\cite{fried2023incoder} as the base model for the initial version of \cc.
Recently, with the release of CodeLlama~\cite{roziere2023code}, we switched to using a fine-tuned version of CodeLlama-7B as the foundation model.

\cc has several desired characteristics for powering an inline suggestion experience: (i) multi-linguality, due to the CodeLlama's training on a multitude of programming languages, (ii) customized for \company, as a result of our internal fine-tuning of CodeLlama on organization-specific data, (iii) natural language proficiency, which gives it the capability to generate and understand inline code comments, and (iv) bi-directionality, due to its fill-in-the-middle (FIM) training that allows it look at both the code before and after the cursor when generating a suggestion.




\subsection{Measures for Evaluating \cc}
\label{sec:method:measures}

To evaluate \cc we used the following measures:

\begin{itemize}
\item \# of suggestions displayed per user per day
\item E2E latency of generating and displaying suggestions
\item Acceptance rate (\# accepted / \# displayed for suggestions shown to user for > 750ms)
\item \# chars accepted per user per day
\item \% keystrokes saved saved (\# chars accepted / \# of chars typed by user)
\end{itemize}

We think of these metrics as a funnel, with suggestions displayed at the top, leading to increased throughput in suggestions accepted, and increasing the total \% of keystrokes saved. Latency serves as a guardrail metric, but also feeds into this funnel since lower latency increases the \# of suggestions displayed.

At \company, changes are rolled out in randomized double blind trials, \ie A/B tests~\cite{NudgeBot2022FSE}. We describe the setup for each of our rollouts to evaluate the effectiveness of each new multi-line feature against the holistic user experience in Section~\ref{sec:production}.

\section{Addressing Challenge 1}
\label{sec:jarring}
\textit{User interface experience - Eliminating the ``jarring'' effect}

In this section, we explore the unique challenges and considerations associated with deploying a coding assistant capable of multi-line suggestions within a large-scale industrial organization like \company. These insights are derived from feedback received from hundreds of users that were using \cc as early adopters. 
Contrary to initial intuition, the multi-line use case presents a higher level of complexity compared to the single-line use case, from both a product perspective and technical implementation standpoint. 
The primary objectives for each use case are as follows: 

\begin{itemize}
\item \textbf{Single-line}: The aim is to provide quick and highly accurate assistance, facilitating task completion and reducing the burden of keystrokes for straightforward and repetitive tasks. 
\item \textbf{Multi-line}: The goal is to offer deeply informed, intelligent assistance that aids users in discovering APIs, best practices, and implementation details.
\end{itemize}

We provide examples of the single-line completion in Figure~\ref{fig:exampleSingle-line-jarring}, the difficulties in multi-line completion in Figure~\ref{fig:exampleMulti-line-jarring}, and our multi-line strategy in Figure~\ref{fig:pre_processing}. The captions for the figures are extensive to allow the reader to walkthrough the examples. 

\subsection{Definition of the ``jarring effect"}

\begin{figure}
\centering
\includegraphics[angle=0, scale=0.55]{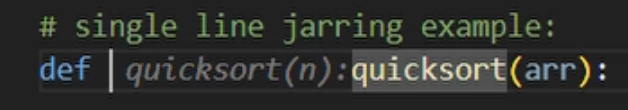}
\caption{Single-line "jarring" effect example: The user cursor positioned between "def" keyword and the "quicksort" function, inline suggestion appears and moves the existing user code to the right. }
\label{fig:exampleSingle-line-jarring}
\end{figure}

\begin{figure}
\centering
\includegraphics[angle=0,width=\imagewidth]{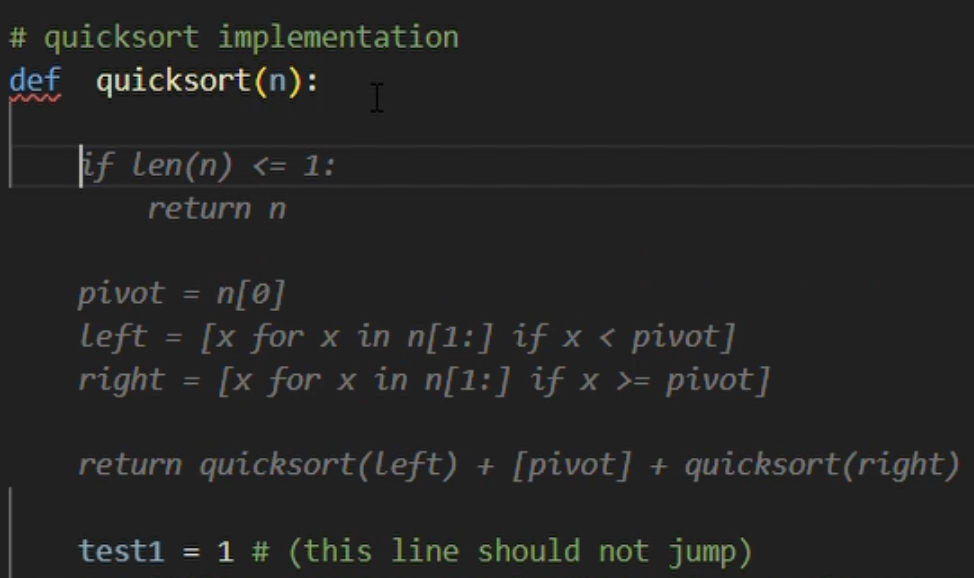}
\caption{Example showing multi-line "jarring" effect: the user cursor was between a function name and the next line containing the statement "test1 = 1". When the suggestion occurs, the existing line is pushed down disrupting the developer's flow and forcing them review the suggested "quicksort" function while also determining the correct location of their existing code. }
\label{fig:exampleMulti-line-jarring}
\end{figure}

\cc suggests code as the user types. This means suggestions appear frequently while the user is engaged in their primary authoring workflow. One of our strictest design principles is that suggestions should not be intrusive and should seamlessly integrate into the user flow. 
Based on user feedback, we identified suggestions that significantly disrupt the user, resulting in a jarring experience. The two primary instances of this are (1) suggestions that shift existing code that the user is currently reading or typing, either upwards, downwards, or to the right as shown in Figure~\ref{fig:exampleSingle-line-jarring} (2) suggestions that do not align with the structure of the code, such as poorly formatted suggestions that overlap the existing scope of the cursor as shown in Figure~\ref{fig:exampleMulti-line-jarring}.

\subsection{Approach to address the ``Jarring Effect"}

Addressing the ``jarring effect" for {\it single-line} suggestions is relatively uncomplicated. To prevent the displacement of existing code, we simply refrain from displaying suggestions if there is any code to the right of the cursor, with certain exceptions such as \}, ) and ]. Since no characters exist on the right-hand side of the cursor when a suggestion appears, no user code can be shifted to the right. Furthermore, as we only suggest until the end of the line, we eliminate the risk of introducing unformatted code that overlaps the existing cursor scope.
However, for multi-line suggestions, the "jarring effect" is more likely to occur, as existing code below the cursor is constantly being pushed up and down.

To circumvent the jarring effect for multi-line suggestions, we adhere to the following rules:
\begin{itemize}
\item Multi-line suggestions are triggered only when the cursor is positioned at the end of the scope that owns the cursor.
\item Multi-line suggestions should be shown until the end of the current scope.
\end{itemize}

When users write code, their flow and mindset is in the most inner scope. Therefore, satisfying rule (1) ensures that when we suggest multi-line to the user, the lines below that are pushed down are in the outer encapsulating scope, thereby not disrupting the user flow and not causing the jarring effect. Furthermore, satisfying rule (2) ensures the suggestion structure is completing until the end of the current scope with no overlap.

We specifically trigger multi-line in the following cases:
\begin{enumerate}
\item The cursor is positioned at the end of the most inner scope that contains it.
\item The cursor is at the end of a line that defines a new symbol that creates a new scope.
\item The cursor is at the end of a notebook cell (for \Bento\cite{bento} notebook use case).
\item The user explicitly requested using a shortcut key. 
\end{enumerate}
\begin{figure}
    \centering
    \begin{subfigure}{\linewidth}
        \centering
        \includegraphics[width=\linewidth]{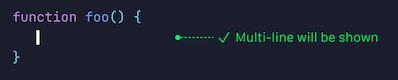}
        \caption{Multi-line suggested since the cursor is positioned at the end of an empty function which is the inner scope.}
    \end{subfigure}
    \hfill
    \begin{subfigure}{\linewidth}
        \centering
        \includegraphics[width=\linewidth]{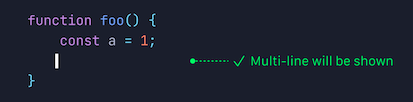}
        \caption{Multi-line suggested since the cursor is positioned at the end of an non-empty function which is the inner scope.}
    \end{subfigure}
    \begin{subfigure}{\linewidth}
        \centering
        \includegraphics[width=\linewidth]{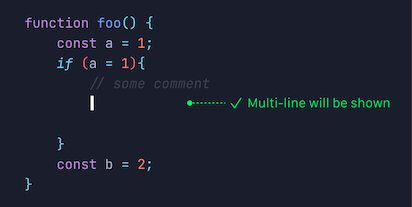}
        \caption{Multi-line suggested since the cursor is at the end of "if statement" which is the inner scope.}
    \end{subfigure}
    \hfill
    \begin{subfigure}{\linewidth}
        \centering
        \includegraphics[width=\linewidth]{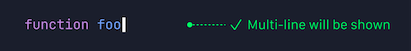}
        \caption{Multi-line suggested since the cursor is at the end of a newly defined function.}
    \end{subfigure}
    \hfill
    \begin{subfigure}{\linewidth}
        \centering
        \includegraphics[width=\linewidth]{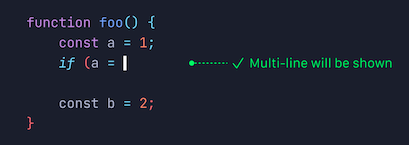}
        \caption{Multi-line suggested since the cursor is at the end of a newly defined "if statement".}
    \end{subfigure}
    \hfill
    \begin{subfigure}{\linewidth}
        \centering
        \includegraphics[width=\linewidth]{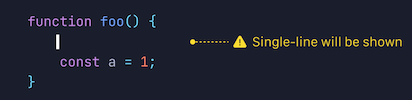}
        \caption{Multi-line is not suggested since the inner scope if function and the cursor is not at the end of it.}
    \end{subfigure}
    \hfill
    \begin{subfigure}{\linewidth}
        \centering
        \includegraphics[width=\linewidth]{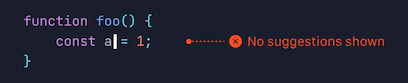}
        \caption{Both multi-line and single-line not suggested since there are characters after the cursor in the same line.}
    \end{subfigure}
    \caption{Examples showing pre-processing stage: Deciding based on the cursor position which type of suggestion should be displayed.}
    \label{fig:pre_processing}
\end{figure}

\subsection{Technical Implementation of the Strategy}
\begin{figure*}
\centering
\fbox{\includegraphics[angle=0,width=2\imagewidth]{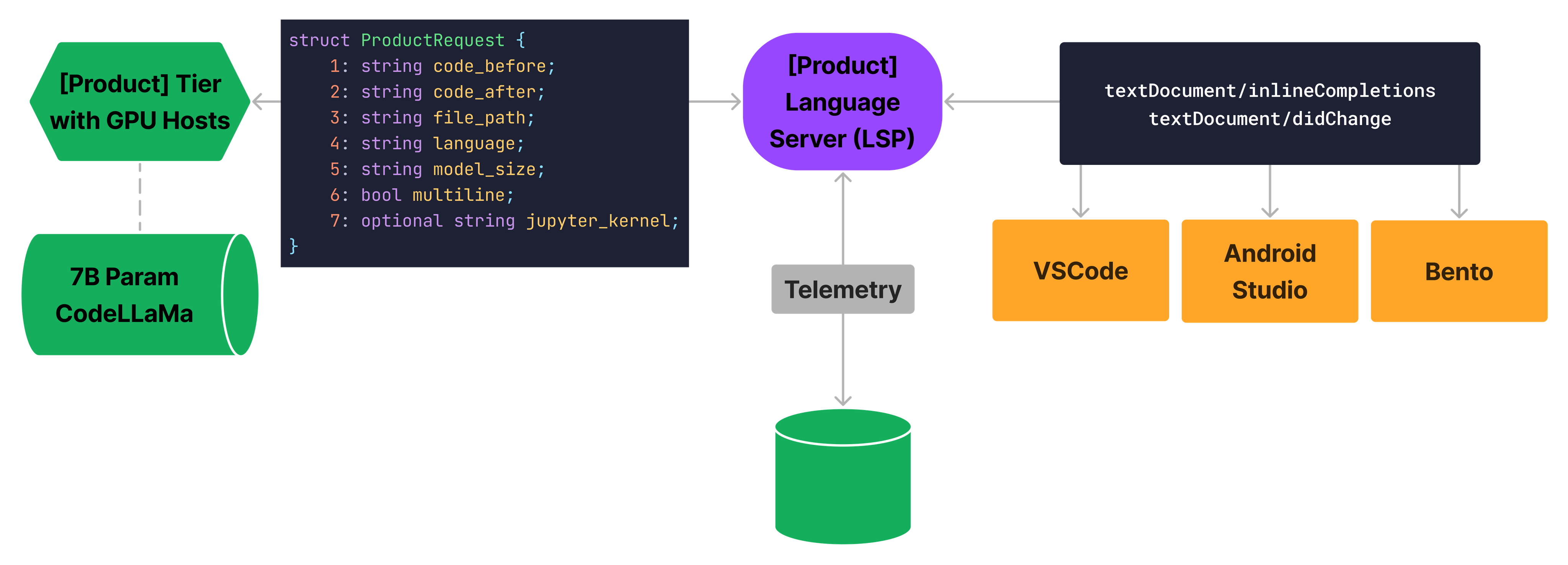}}
\caption{System architecture of \cc: Client editor that surface the suggestions, a language server to mediate requests with \cc model service host. In the request "multi-line" flag is passed to the model service. }
\label{fig:arch}
\end{figure*}

The requirement to trigger multi-line only in certain cases, based on the cursor position, necessitated the integration of semantic context understanding into the \cc system.
The existing \cc system employs a typical client-server architecture (shown in Figure~\ref{fig:arch}) in which the server is an inference tier that runs the model and a client is an editor that surfaces code suggestions. We encode the bulk of the client-side logic in a Language Server Protocol (LSP) conformant language server that is reused across multiple editor integrations. To mediate requests between the client and server, we implemented a language server in Rust that we reuse across our various editor integration. While most language servers are designed to support a wide array of traditional IDE functionality (autocomplete, jump-to-definition, etc.), the \cc language server supports only one meaningful request type: "textDocument \slash inlineCompletions".

Avoiding the jarring effect for single-line was trivial from a technical point of view as it required a simple conditional statement. For multi-line on the other hand, the strategy described above is much more complex and requires detailed semantic context understanding of the programming language. 

\textit{Request workflow - Pre-processing}:
A completion request is sent from the client to the language server, the \cc language server parses the current state of the file and understands the cursor position. If the cursor's position aligns with one of the cases outlined above, this request is marked as a multi-line request. The request is then validated in the local cache; if it is a cache miss, it is sent to the \cc model flagged as a multi-line request (examples shown in Figure~\ref{fig:pre_processing}). 

\begin{figure}
\centering
\includegraphics[angle=0,scale=0.5]{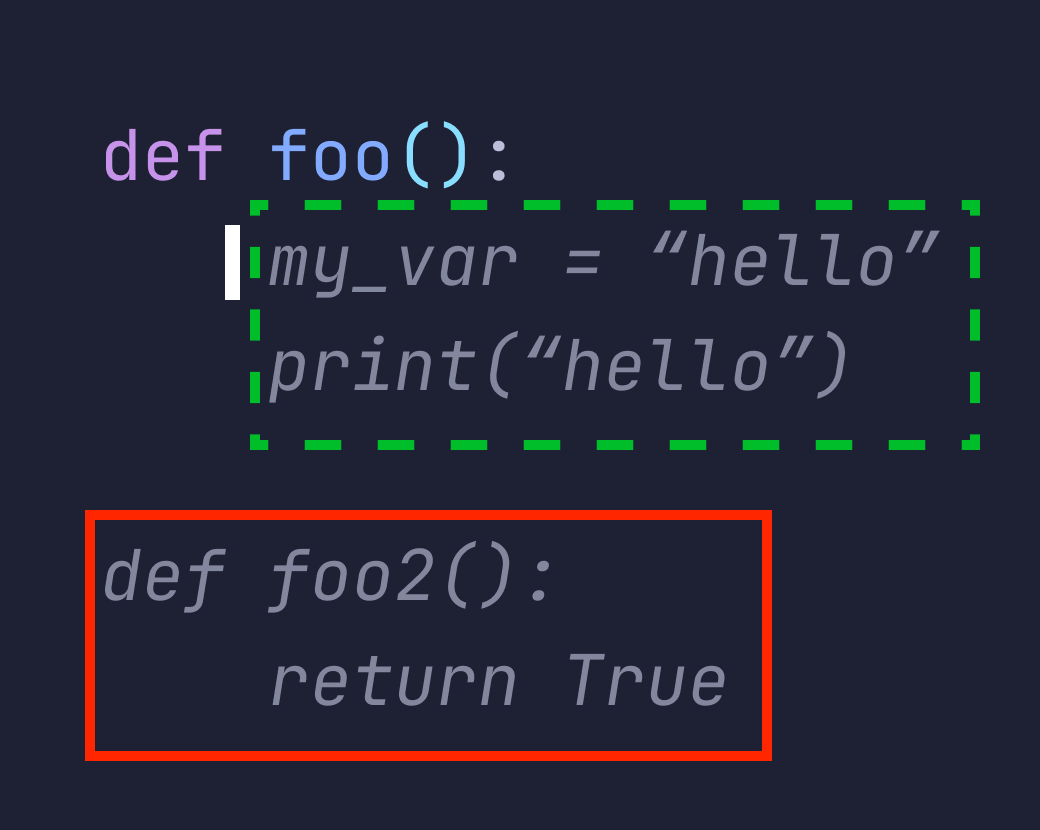}
\caption{Example showing the post-processing stage: The cursor is in the scope of the ``foo'' function. Although, the model returns a multi-line suggestion of both the ``foo" and "foo2" functions, postprocessing will remove the code in the red box, and will only display suggestions for the in-scope ``foo'' function to the user.}
\label{fig:exampleMulti-line_truncation}
\end{figure}

\textit{Request workflow - Post-processing}:
The model LLM generates suggestions and sends back the response to the language server. The model response does not have a strict guarantee for how well it will be structured; it may contain overlap with code in the current scope. Therefore, in the language server, the response goes through an additional step of post-processing that truncates the response in case it overlaps the current scope, as shown in Figure~\ref{fig:exampleMulti-line_truncation}.

Finally, we built an extensive unit test suite (473 tests) to validate that there was no end-to-end jarring effect or incorrect truncation with our implementation. 

\begin{tcolorbox}
In summary: (1) Multi-line suggestions are triggered only when the cursor is at the end of scope (2) Suggestions are shown until the end of the current block (3) After the suggestion is accepted, the cursor needs to be moved to the end of the suggested block.
\end{tcolorbox}

\section{Addressing Challenge 2}
\label{sec:model}
\textit{Responsive User Experience}

\cc's inline suggestions show up as the user types, each suggestion aligns with the given state of the file and invalidated (dismissed in the editor's User Interface) as soon as the user types an additional keystroke or moves their cursor. This implies that latency is a key factor in determining the "display rate" - the percentage of suggestions shown to users. 

For single-line we limit the generation of a suggestion to stop at either newline or 25 max tokens. The average latency is low, therefore display rate is not a major concern. Multi-line generations are much longer (p50 325 chars, p90 450 chars), and max tokens is set to 120. This can take over 2 seconds for large suggestions; therefore, reducing latency for multi-line is significantly more likely to increase the display rate. 
In this section we describe multiple approaches and projects that were built focused on reducing the latency of generating long multi-line generations and increasing its display rate. The improvements were done across all main components of the \cc System - the client-side editor extension, the language server and the service hosting the model.

\subsection{Improvements in the editor client extension (i.e. VSCode / \Bento\cite{bento} Notebooks)} 
During initial testing, \cc's multi-line suggestions were generated automatically as the user typed. However, from the user's perspective, there was not a deterministic way to know whether a \cc suggestion was still being processed and would appear momentarily, or whether \cc had decided not to display a suggestion. This unpredictability caused frustration to the user and made the system appear unreliable.

\begin{figure}
\includegraphics[angle=0,scale=0.6]{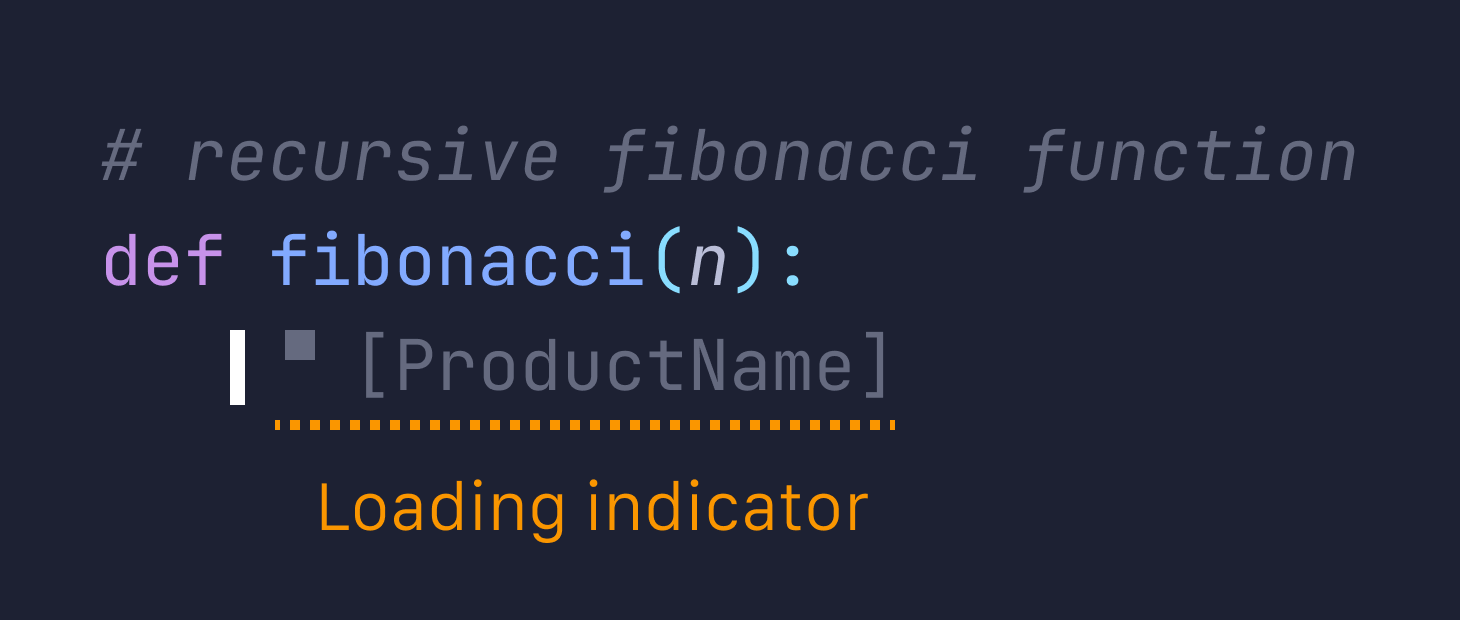}
\caption{Inline indicator (marked with orange underline) with a spinner shown to the left of the cursor alerting the user \cc may show soon an AI suggestion}.
\label{fig:inline-indicator}
\end{figure}

To solve this, we introduced an inline spinner indicator, as seen in Figure~\ref{fig:inline-indicator}, which pops up next to the user cursor as long as there is an active completion that still needs to be responded to. As soon as the language server's pre-processing (described in \ref{sec:jarring}) determines that the current request should be multi-line, it sends a "\cc/fetchingMultiline" notification to the editor client, which then renders the indicator. The editor dismisses the indicator once the request is responded to or canceled.

The user's mental model is rooted in knowing whether a suggestion is expected to appear or not. The inline indicator lets users know if a multi-line suggestion is being processed, and therefore allows them to make a decision on whether to type the next character -- or to wait a little longer to see the upcoming suggestion. This feedback improves the overall display rate by nudging users to wait a little longer, especially for longer multi-line suggestions. This results in a transparent, predictable experience for users that helps build their trust in the \cc system. 

\begin{tcolorbox}

Because multi-line suggestions take significantly longer to generate, we added a spinning UI indicator so that users were aware that a multi-line suggestion was being generated. This reduced the perceived latency and increased usage of multi-line suggestions over single-line suggestions.
    
\end{tcolorbox}

\subsection{Optimizations to the model-hosting service}

We implemented numerous improvements to the model-hosting service with the primary goal of reducing latency. Our hypothesis was that latency reduction would increase all of our quantitative outcome metrics, particularly for multiline suggestions where generation time was significantly longer. Suggestions being available sooner would mean that more suggestions were shown to users, before they would have taken another action (e.g. typing the next character) that would cause the suggestion not be shown.

From mid-June through September, we ran a series of A/B experiments \cite{NudgeBot2022FSE} on \cc, which is rolled out to 10's of thousands of \company's engineers, that incrementally improve latency: We initially rolled out to 10\% of the population as a smoke test, before comparing metrics at 50\% rollout for a one week duration to validate the latency impact across all our outcome metrics. Below, we highlight each of the latency experiments we ran, then summarize the net metric gains in Section \ref{sec:production}.

{\bf Flash Attention.} When we implemented Flash Attention \cite{dao2022flashattention} on our 6.7B InCoder-based FIM-trained model on multi-line suggestions, we saw a median 2.53x reduction in first token latency and median 1.15x reduction in latency to the final token. Flash Attention is an improvement primarily in the model's attention stack, so a greater improvement in first token latency was expected. In our experiment, we found a 13\% increase in the number of suggestions displayed, an 8\% increase in acceptance rate, a 3\% increase in the average length of suggestions accepted. Overall, we saw a 74\% reduction in median latency because Flash Attention improved single-line suggestion latency even more than multi-line and single-line suggestions are the majority of suggestions that our models generate.

{\bf CUDA graphs and fused kernels.} Using CUDA graphs and fused kernels\cite{xformers} together on our 6.7B InCoder-based FIM-trained model, we observed that a reduction in our median generation latency by 29\%, and at the 90th percentile a reduction in latency of 40\%. The latency reductions led to increases in our absolute number of displayed suggestions by 26\% and total number of suggestions accepted by 17\%. However, we also observed the acceptance rate decline of 1\%. From this experiment, we learned that acceptance rate is not strictly correlated with latency. We hypothesize that the improvement in latency during this experiment resulted in suggestions being shown in environments of increasingly low acceptance rates. For example, if the user rejected a suggestion then typed only one more character, a newly generated suggestion was unlikely to be accepted even if displayed immediately with low latency.


{\bf Queue Priority.} We studied the effect of the longer multi-line suggestions on our GPU capacity utilization. We discovered that longer generations have an outsized effect on capacity load. Additionally, requests for longer multi-line suggestions were receiving time-out responses at nearly 2x the rate of single-line requests. We suspect the reason for the large difference between time-out rates is created by the triggering strategies we employ (as described in Section~\ref{sec:jarring}). To alleviate the queue contention, we implemented QoS to advantage our longer multi-line requests. We increased the queue gestation time relative to shorter single-line requests to boost the multi-line success rate. This effort succeeded in reducing the magnitude of multi-line time-out responses. We succeeded in increasing the number of displayed multi-line suggestions and observed a 1\% absolute increase in keystrokes saved. 

{\bf Streaming and early cancellation.} Building on our work to truncate generated suggestions at the end of scope (see Section 3.3), our analysis discovered that 54\% of all characters generated were being truncated before being shown. Further, 47\% of all suggestions generated by the model were never displayed. To reduce this wasted effort by our GPU capacity, we introduced a streaming response mechanism that allowed early cancellation when the client closes the stream. With streaming, the client is able to communicate back to the model service that either the generated suggestion is no longer needed at all or that the model has generated characters that will be truncated. We observed a 45\% improvement to overall model service request round trip latency as a result of implementing streaming with cancellation.

{\bf Batching.} We used 2-way tensor parallelism to serve responses 1.5x faster. Since this uses two GPUs to serve a request, but does not improve the latency by a factor of two, it effectively requires more capacity to handle the same workload. However, we found that this can be more than compensated by using continuous batching, giving us significantly faster responses while also gaining improved effective capacity due to the efficacy of batching multiple requests.

Improving latency has a compounding effect. Reducing latency distributions increases the effective capacity of our machines to serve more requests. For instance, the above optimizations we can serve the same request with a 2.5x reduction in the number of GPUs. The main reason is that there is less wasted computation on cancelled requests and request can be batched.  Reducing capacity utilization further reduces queuing times and the distribution of round trip latency.

By combining streaming alongside tensor parallelism and continuous batching, we sped up the median singleline suggestion from 440ms to 280ms (1.5x faster/35\% latency reduction), and the median multi-line suggestion from 2000ms to 750ms (2.5x faster/60\% latency reduction). This in turn led to a 16\% relative improvement in characters accepted by users. This confirms that latency has a significant impact on the effectiveness of code completion tools.

\begin{tcolorbox}
We invested in five techniques to optimize the model-hosting service, which reduced median latency of multi-line suggestions down to 750ms. This in turn increased the \# of suggestions shown to users and the overall keystrokes saved by users, as validated through A/B experiments in Section \ref{sec:production}. Having a responsive UX was crucial to the adoption of multi-line suggestions.
\end{tcolorbox}

\section{Addressing Challenge 3}
\label{sec:production}
\textit{Production release and effectiveness measurement}

The final challenge faced was in monitoring the effectiveness of the experimental features enumerated in the prior sections. We tracked a host of metrics in real-time to evaluate whether each feature was improving the user experience, and adjusted the features accordingly. 
These metrics included acceptance rate, display rate, latency, \% keystrokes saved, and \# of characters accepted per user. The metrics were broken down by single-line vs multi-line suggestions to determine the net benefit of each feature for users. We ran A/B experiments with a 50/50 split to control and test for each feature release.

\subsection{Experiments for Release of Multi-line Suggestions}
\label{sec:production:metrics}

First, when rolling out each of the multi-line trigger points (as explained in Figure~\ref{fig:pre_processing}), we closely monitored both acceptance rate and throughput for any negative impact on the user experience:

\begin{enumerate}
\item Multi-line suggestions require more mental effort for users, as they must review multiple lines rather than reading a single-line suggestion
\item Multi-line suggestions cause a substitution effect -- some of the trigger points (e.g. upon hitting newline in a new function block) overlap with single-line. Due to the longer latency for generating multi-line suggestions, users may end up typing past these trigger points, causing a missed display opportunity for \cc to save the users on typed keystrokes (whereas a single-line suggestion would have displayed, given its lower latency)
\item Users may sometimes prefer seeing consecutive single-line suggestions, instead of one or two longer multi-line suggestions, depending on their workflows
\end{enumerate}
 
\begin{figure}
\includegraphics[angle=0,width=\imagewidth]{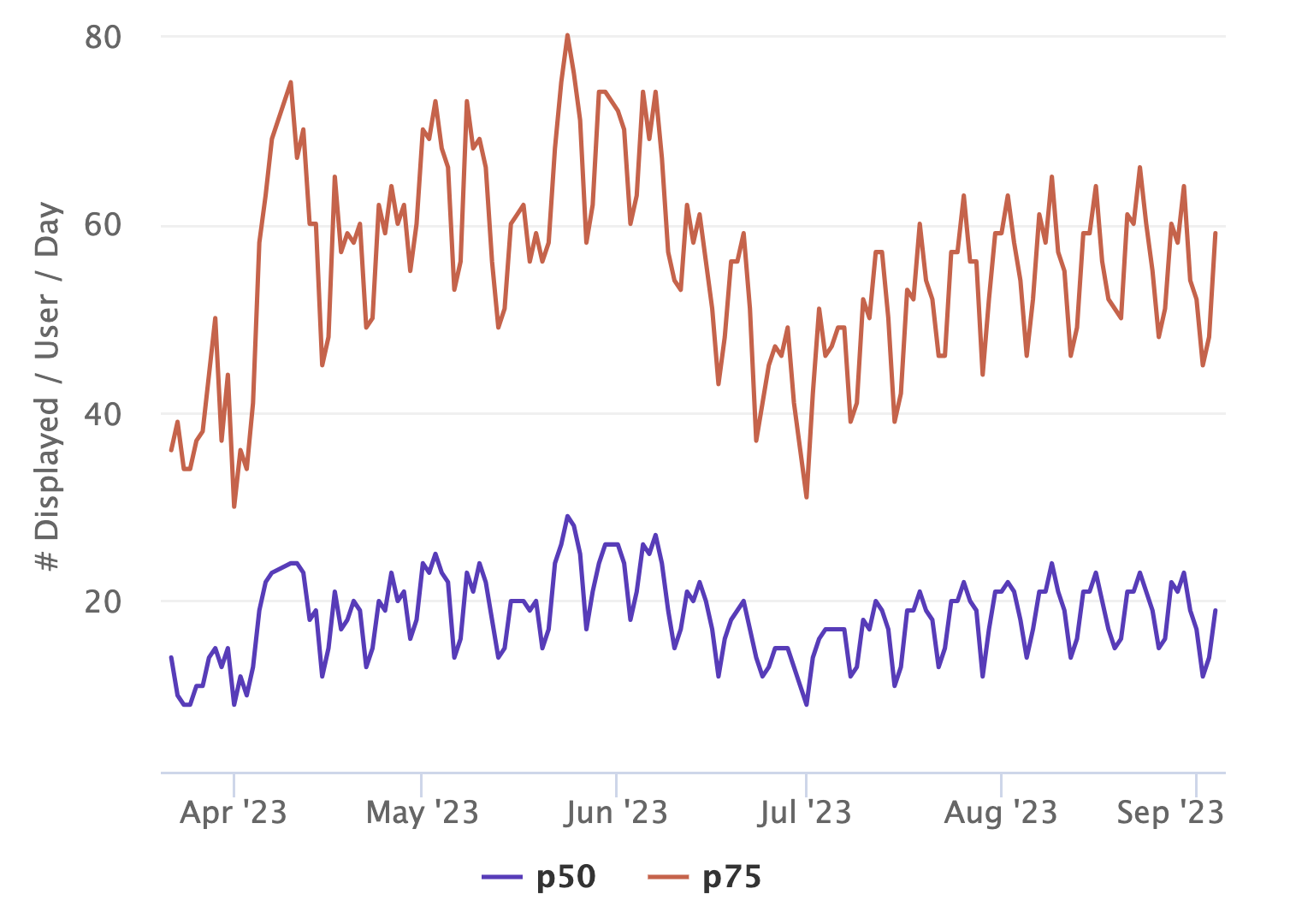}
\caption{The \# of displayed suggestions per user per day, at p50 and p75. This display metric dropped in mid-June as higher latency multi-line suggestions rolled out, but recovered by end of August due to our investments in perceived latency.}
\label{fig:rollout-displayed}
\end{figure}

As seen in Figure~\ref{fig:rollout-displayed}, when initially rolling out multi-line suggestions in mid-June, the number of displayed suggestions per user dropped significantly, due to the longer generation time which led to users typing past the trigger point. 

 However, due to our investments in (1) reducing median multi-line latency from 2000ms to 750ms (2) showing the inline indicator (both of which were described in Section~\ref{sec:model}), this display metric gradually restored close to original levels by end of August. These latency investments decreased the gap in perceived latency between single-line and multi-line suggestions, which in turn contributed to the continued increase in throughput as users found multi-line suggestions more and more valuable with the increased responsiveness of the suggestion UI.

Through our production monitoring, we found that acceptance rate also remained similar at 29\% between the two suggestion types, reinforcing that despite the longer suggestions (which are harder to predict accurately), multi-line suggestions were still favored at similar rates as single-line. Hence overall throughput increases since more lines of codes were accepted at a given display opportunity.

\begin{figure}
\includegraphics[angle=0,width=\imagewidth]{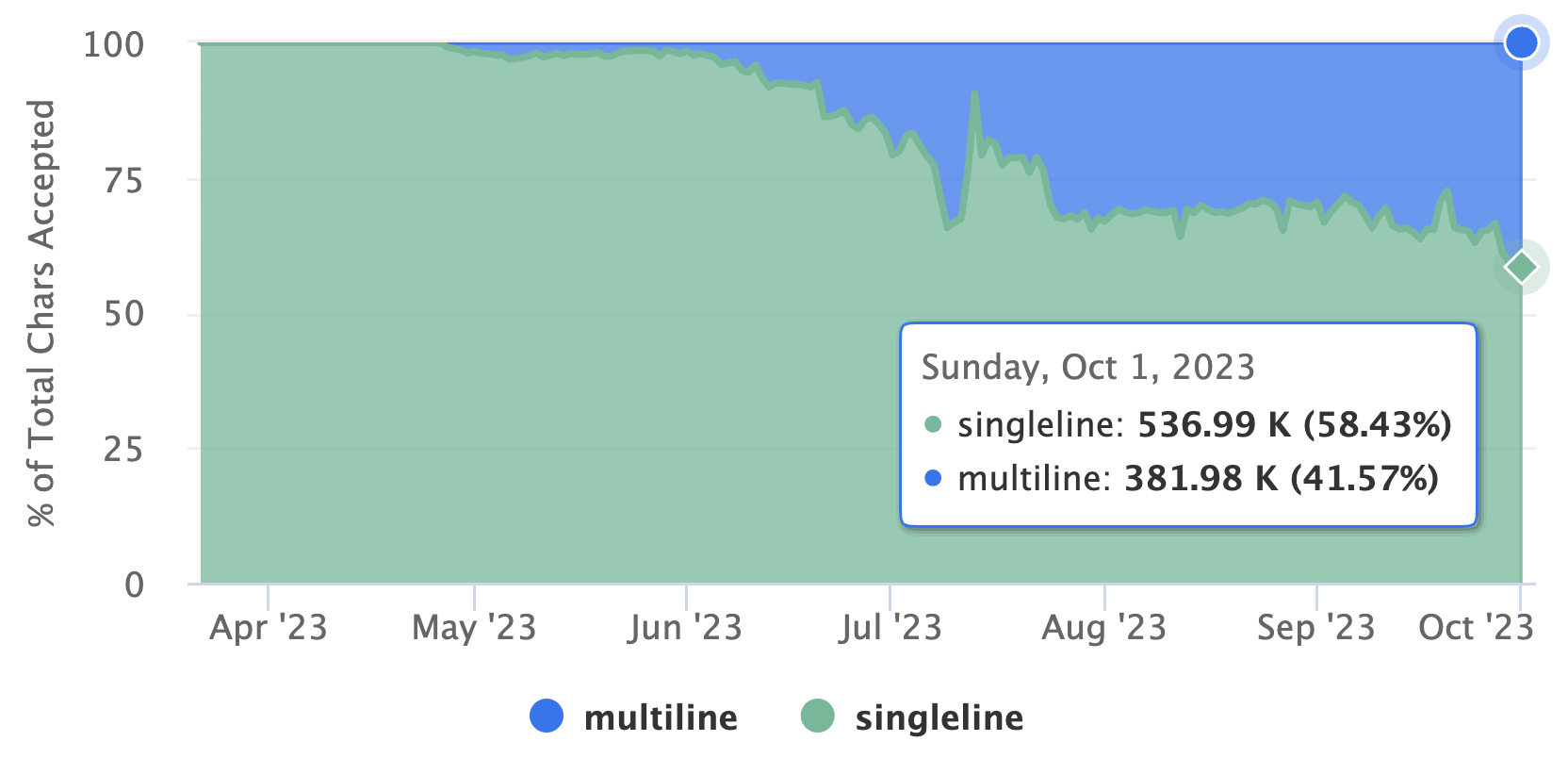}
\caption{Total characters accepted, broken down by suggestion type. Multi-line suggestions trigger with higher latency and lower frequency (16\% of total volume of displayed suggestions). Yet multi-line suggestions account for 42\% of total characters accepted for users.}
\label{fig:chars-accepted-comparison}
\end{figure}

Furthermore, since we triggered multi-line suggestions less frequently to reduce the jarring effect, multi-line suggestions only accounted for roughly 16\% of total volume of suggestions displayed. Yet as shown in Figure~\ref{fig:chars-accepted-comparison}, multi-line suggestions accounted for 42\% of total characters accepted by users. These throughput metrics demonstrated the success of our investments in reducing user perceived latency, which in turn increased usage of multi-line suggestions relative to single-line suggestions.

Finally, our online monitoring metrics demonstrate that multi-line suggestions are the correct surface for further improvement to the product over time. As shown in Table~\ref{fig:keystrokes}, when releasing the CodeLlama-7B model, we observed that the impact on multi-line was much greater: +40\% relative improvement in multi-line throughput, compared to only +25\% relative improvement in single-line throughput. This opens the door to further investments in generating longer and longer suggestions -- for example, by increasing the max token length for generation, or by triggering multi-line suggestions at more locations.

\begin{table}
\includegraphics[angle=0,width=\linewidth]{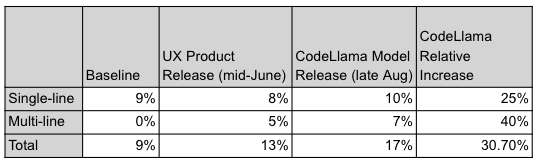}
\caption{The \% keystrokes saved at each step of the release cycle. (a) The combined effect from the UX Product Release (mid-June to August) was a 5\% absolute increase from multi-line, while taking away only 1\% absolute from single-line, demonstrating a net improvement to the user of 4\% absolute. (b) A new model release (CodeLlama-7B) further showed the value of multi-line as the improvement surface going forward, as the new model increased single-line throughput from 8\% to 10\% (+25\% relative), but a much larger increase of 5\% to 7\% (+40\% relative) for multi-line.}
\label{fig:keystrokes}
\end{table}

\subsection{User Feedback and Opt-out Rate}
\label{sec:production:user}

We monitored qualitative user feedback to triangulate the metric improvements we observed in the prior section. Before \cc handled the "jarring" effect, and before new CodeLlama-7B was deployed, several users complained about the "jarring effect" and quality of multi-line suggestions:
\begin{itemize}
\item \textit{"[Negative] Lately, I've been finding Python multi-line suggestions distracting. This DevX feels substantially different than single-line completions:
It shifts my code around, causing me to shift my gaze from the line I'm writing to figure out what happened.
Sometimes the code is useful, but it's largely been not useful compared to single-line completions (I imagine we see this in suggestion accept rate data).
I wonder how we can make this DevX less interruptive, while we work on improving suggestion quality."}
\end{itemize}

After developing \cc's multi-line algorithm, the feedback was much more positive:
\begin{itemize}
\item \textit{"[Positive] I had so many moments of surprise and delight to the point of taking screenshots of suggestions. \cc beginning to automate so much of the boring/painful parts of coding to make it super fun \& fast. I haven't felt a magical productivity leap like this since <infra feature change from several years ago>."}
\item \textit{"[Positive] I had no idea <language> had multi-line strings; this popped up when in a situation where I was better off using one and was grateful for the discovery"}
\item \textit{"[Positive] Regarding \cc: Have I become that predictable as a human being!
My intention was to write a Python function that generates a SQL case-when statement based on a dictionary I previously defined. I typed the name of the function and it autogenerated the 5-line return statement. It correctly guessed I wanted to create a SQL case-when statement and wrote the function just like I was thinking in my head!!! I always thought the way an individual writes code is unique to that individual! So have I become that predictable as a human being! What even is real, anymore!"}
\item \textit{"[Positive] Dealing with annoying boilerplate, ex. propagating 5 variables through state -> query -> response. \cc makes faster work of these than my usual copy+paste+select+change motion [...] Both writing the generic function \& replacing the code was 50\%+ accepted suggestions"} 
\item \textit{"[Positive] \cc is just amazing when writing generic functions like this. Of course, this isn't perfect since it did not get the else part of the function correct but still helped me reduce my keystrokes by 90\%."}
\item \textit{"[Mixed] I've been noticing several cases where the first lines of the \cc suggestion are quite good, but the suggestion gets a bit too ambitious and goes on several lines beyond that with code that is subjectively not what I want and objectively less related to the intent I expressed with what I had typed so far."}
\item \textit{"[Negative] I find the multi-line suggestions are rarely useful, and cause issues if I accidentally accept them."}
\end{itemize}

User feedback underscores the intricate equilibrium required when launching a feature like multi-line suggestions. This balance is between recall -- as manifested in the display rate and the number of lines in each suggestion -- and the quality of the suggestion itself. As we enhanced the quality of multi-line suggestions by mitigating any jarring effects and improving the model's quality, we observed a corresponding improvement in user feedback. While we leave a full thematic or ground theory analysis of the qualitative feedback as future work, the initial user feedback has been overwhelmingly positive, as shown in the above testimonials. 

We also note that very few engineers disabled the multi-line feature. The overall opt-out rate of \cc stayed constant over time at $< 0.8\%$ of all users. We allowed users to specifically toggle off multi-line suggestions, while keeping single-line suggestions on. The multi-line opt-out rate here is even lower at $< 0.1\%$ of all users, indicating widespread adoption and favorability toward multi-line suggestions.


\begin{tcolorbox}
During the roll-out of multi-line suggestions, we monitored metrics like acceptance rate, display rate, latency, and throughput to evaluate the net benefit of multi-line suggestions vs. single-line suggestions. These metrics demonstrated that the investment in multi-line suggestions disproportionately increased throughput, as multi-line suggestions accounted for 42\% of total characters accepted per user per day (despite only accounting for at 16\% of displayed suggestions), and drove the significant net increase in percentage of keystrokes saved nearly doubling from 9\% to 17\% (as shown in Table~\ref{fig:keystrokes}). Less than $1\%$ of engineers opted out and disabled multi-line suggestions.
\end{tcolorbox}

\section{Threats to Validity}
\label{sec:threats}

\subsection{Generalizability}
This study is conducted at \Meta. Although this is a single company it includes codebases many times larger than those typically studied in open source projects. It also contains code that covers a wide range of software systems from social network to infrastructure projects, such as databases and continuous integration systems. There is also software that runs directly on VR hardware. We are able to release \cc to 10's of thousands of engineers. Our results may not generalize to smaller companies and projects that do not have a large codebase to train their models on.

\subsection{Construct and Internal Validity}

In Section~\ref{sec:method:measures}, we describe the metrics we track to determine if multi-line \cc is effective. There have been few large scale deployments of AI-assisted, so we used the existing infrastructure at \Meta to track: display rate, latency, \% keystrokes saved, \# of characters accepted per user, and opt-out rate. In our A/B tests across 10's of thousands of engineers, the results were positive for each metric and convincing enough to allow us to roll out multi-line to the entire company. We created dashboards to continue to monitor these metrics. In future work, we would like to track metrics that other researchers and companies have found useful.

\section{Discussion and Related Work}
\label{sec:literatureAndDiscussion}

While there is a large body of software engineering research work on code completions \cite{bruch2009learning,robles2008program,proksch2015intelligent,zhou22improving,kim21code}, there has been limited deployment of these in large industrial environments \cite{intellicode,codewhisperer,copilot,googleblog}. With the advent of Generative AI, advances in the code assistance area are now in weeks rather than months. Several of these results are circulated as blog posts and not traditional papers, with limited discussion of evaluating the trade-offs between single-line and multi-line suggestions. 

Hence, while multi-line support is common for AI code assistants, to the best of our knowledge, we are the first to describe contributions in:
\begin{enumerate}
\item Using scope-based multi-line algorithm for AI-assisted suggestions
\item Applying general LLM optimizations to improve the responsiveness of AI-assisted suggestions in an enterprise product
\item Deploying A/B industrial-scale rollouts to quantify the effectiveness of multi-line vs single-line suggestions
\end{enumerate}

\subsection{Scope-Based Multi-Line Algorithm}

While many tools provide multi-line suggestions, we did not find any discussion of the algorithm that they use to avoid the ``jarring'' effect or to thoughtfully limit the amount of code shown. For example, Amazon's CodeWhisperer \cite{codewhisperer} can be configured to produce multi-line suggestions, but the suggestions appear to run unlimited beyond the current semantic scope, causing a constant distraction for the user. They also seem to be preferentially triggered in highly specific contexts like a docstring, rather than through an automatic semantic-based algorithm as described in Section~\ref{sec:jarring}.

Google's ML code completion\cite{googleblog} does automatically constrain their multi-line suggestions based on a semantic filter, but this results in a far more restrictive algorithm compared to our approach. They show both the next token and the full line suggestion in a dropdown menu, with the length of the completion  dependent on the current token. While the authors show a multi-line completion, it is unclear how many additional lines or what other restrictions are placed upon the completions, and no multi-line algorithm is described. 

In contract, \cc's multi-line algorithm automatically triggers as the user types -- while remaining judicious in which trigger points are used, and selectively limiting the generated suggestion to the user's current scope. Multi-line suggestions are more difficult to accurately generate, but the scope-based algorithm enables us to display suggestions that fit within the user's current thinking context, thus supporting their train of thought without creating additional noise.

\subsection{Responsive UX}
With regards to latency reductions to make the product more responsive, many of the techniques we describe are standard in software engineering (e.g. early cancellation / streaming), or published techniques for optimizing transformer performance (e.g. Flash Attention\cite{dao2022flashattention}, xFormers\cite{xformers}). However there is little published research on the impact of deploying these in an industrial setting. While various external product releases, such as Copilot, Codeium, CodeWhisperer \cite{CodeiumAmazonComparison, CoPilotLatency}, mention latency reductions as important drivers for improving the quality of the user experience, they do not offer a rigorous evaluation of latency impact across a suite of product metrics. 

Hence, to the best of our knowledge, we are the first to contribute both an enumeration of the specific algorithms that drive these latency reductions, and a measurement of the impact of these when released in a production environment.

\subsection{Evaluating Effectiveness of Multi-Line vs Single-Line Suggestions}
As described in Section~\ref{sec:production}, it is not straightforward to determine whether multi-line suggestions are a net improvement to the product experience, as a series of rapid single-line suggestions could be received better by users. To the best of our knowledge, we are the first to show industry metrics from A/B rollouts that demonstrate the trade-off between single-line and multi-line suggestions. As Section~\ref{sec:production:metrics} showed, our experiment metrics provide evidence for the hypothesis that despite latency tradeoffs, longer multi-line suggestions are more impactful to users than a series of shorter single-line suggestions.

\textbf{Quantitative Comparisons.}
Although a direct comparison of quantitative results is impossible between products (given the different contexts, users, and tooling), when similar measures are reported, we can compare the overall trends. At Google, engineers \cite{googleblog} deployed a hybrid semantic ML code completion to 10k+ Google employees (over three months across eight programming languages). They observed a 6\% reduction in coding iteration time (time between builds and tests) for users exposed to single-line ML completion compared to a control group. The authors measured a relatively low keystrokes savings: 0.6\% saved from multi-line, compared to 7\% of multi-line keystroke savings in \cc. This differences indicates either lower display rate or lower number of lines in each suggestion generated by Google. The authors mention that a semantic filtering is applied during post-processing phase -- this type of filtering is not applied in \cc, and may explain a lower display rate in Google's product.

In contrast, at \Meta, rather than using a semantic filter and dropdown menu constraint, \cc completions are shown directly in the file, and are dependent on the position of the cursor. Hence, our suggestions automatically trigger in more cursor locations, likely leading to a higher reported display rate and overall throughput.

Amazon's CodeWhisperer \cite{codewhisperer} is a fully functional code completion tool integrated into the IDE. Amazon's analysis found that language models can generate code with correct syntax and pass unit tests in programming languages they are not intentionally trained on. The tool can be configured to produce multi-line suggestions, but these suggestions appear to trigger in more limited contexts and with significant latency for the user, making it unclear how effective multi-line is. We were unable to find any details from Amazon on the quantitative metric impact that single-line vs multi-line suggestions have on the overall product.

There have been several empirical evaluations of GitHub’s Copilot \cite{copilot} in actual use for automatic code completion. Nguyen et al. \cite{nguyen2022empirical} used 33 LeetCode questions to create queries for Copilot in four different programming languages. They found that Copilot's Java suggestions have the highest correctness score (57\%) while JavaScript is the lowest (27\%). Overall, Copilot's suggestions had low complexity with no notable differences between the programming languages. They also observed some potential Copilot shortcomings: such as generating code that can be further simplified and code that relies on undefined helper methods \cite{nguyen2022empirical}.

\textbf{Qualitative Comparisons.}
Not all industry usage of code completions has been positive. A user study \cite{vaithilingam2022expectation} with 24 participants to understand how programmers use and perceive Copilot found that while it did not necessarily improve the task completion time or success rate, most participants preferred to use it in daily programming tasks, since Copilot often provided a useful starting point and saved the effort of searching online. However, participants did face difficulties in understanding, editing, and debugging code snippets generated by Copilot, which significantly hindered their task-solving effectiveness \cite{vaithilingam2022expectation}. 

These mixed results were also echoed by Bird \etal~\cite{Bird2023Queue}, that found while developer perceived an acceleration in coding, they often had to spend much more time reviewing the generated code. On the other hand, some other studies found that generated suggestions helped in discovering new APIs~\cite{Barke2023OOPSLA1}. 

Our prior work on CodeCompose found an overwhelming 91.5\% positive feedback for single-line completions accelerating the coding and allowing developers to discover new APIs across the monorepo~\cite{codecompose2024}. Likewise, the user feedback described in Section~\ref{sec:production:user} shows that users found a noticeable improvement in their experience as multi-line suggestions rolled out, indicating success with generating longer blocks of boilerplate code and more complex suggestions that help with discovery. Thus our work lays a foundation for the broader community to continue investing in this direction with generating multi-line suggestions while maintaining a positive user experience.

\section{Conclusion and Contribution}
\label{sec:conclusion}

In this paper, we make the following contributions related to the challenges we addressed: 

\begin{itemize}

\item \textbf{Challenge 1. The Jarring Effect}: We developed the following scope-based algorithm (1) Multi-line suggestions are triggered only when the cursor is at the end of scope (2) Suggestions are shown until the end of the current block (3) After the suggestion is accepted, the cursor needs to be moved to the end of the suggested block.

\item \textbf{Challenge 2. Responsive UX}: Multi-line suggestions take significantly longer to generate; hence we had to reduce the perceived user latency to ensure user adoption compared to single-line suggestions. This was done by (i) adding a UI indicator so users were aware of that a multi-line suggestion as being generated (ii) running optimizations to the model hosting service (e.g. Flash Attention, persistent K-V cache)

\item \textbf{Challenge 3. Production Release Effectiveness}: During the roll-out of multi-line suggestions, we monitored metrics like acceptance rate, display rate, latency, and throughput to evaluate the net benefit of multi-line suggestions vs. single-line suggestions. These metrics demonstrated that the investment in multi-line suggestions disproportionately increased throughput, as multi-line suggestions accounted for 42\% of total characters accepted per user per day (despite only accounting for at 16\% of displayed suggestions), and drove the significant net increase in percentage of keystrokes saved nearly doubling from 9\% to 17\%. Less than $1\%$ of engineers opted out and disabled multi-line suggestions.

\end{itemize}

\section{Acknowledgements} 
We want to acknowledge and thank the following people for their work, help, and support in building the multi-line \cc experience at \company: Dian Belanger, Michael Bolin, Renuka Fernandez, Negar Ghorbani, Kelly Hirano, Diana Hsu, Kristian Kristensen, Killian Murphy, Chris Nixon, Zach Rait, Marshall Roch, Shahin Sefati, and Yiru Zhu.

\balance
\bibliographystyle{ACM-Reference-Format}
\bibliography{ref.bib}


\begin{thebibliography}{22}


\ifx \showCODEN    \undefined \def \showCODEN     #1{\unskip}     \fi
\ifx \showDOI      \undefined \def \showDOI       #1{#1}\fi
\ifx \showISBNx    \undefined \def \showISBNx     #1{\unskip}     \fi
\ifx \showISBNxiii \undefined \def \showISBNxiii  #1{\unskip}     \fi
\ifx \showISSN     \undefined \def \showISSN      #1{\unskip}     \fi
\ifx \showLCCN     \undefined \def \showLCCN      #1{\unskip}     \fi
\ifx \shownote     \undefined \def \shownote      #1{#1}          \fi
\ifx \showarticletitle \undefined \def \showarticletitle #1{#1}   \fi
\ifx \showURL      \undefined \def \showURL       {\relax}        \fi
\providecommand\bibfield[2]{#2}
\providecommand\bibinfo[2]{#2}
\providecommand\natexlab[1]{#1}
\providecommand\showeprint[2][]{arXiv:#2}

\bibitem[ben(2021)]%
        {bento}
Meta \bibinfo{year}{Accessed 2021}\natexlab{}.
\newblock \bibinfo{booktitle}{\emph{ELI5: Bento - Interactive Notebook that
  Empowers Development Collaboration and Best Practices}}.
\newblock Meta.
\newblock
\urldef\tempurl%
\url{https://developers.facebook.com/blog/post/2021/09/20/eli5-bento-interactive-notebook-empowers-development-collaboration-best-practices/}
\showURL{%
\tempurl}


\bibitem[cop(2021)]%
        {copilot}
GitHub \bibinfo{year}{Accessed 2021}\natexlab{}.
\newblock \bibinfo{booktitle}{\emph{GitHub Copilot}}.
\newblock GitHub.
\newblock
\urldef\tempurl%
\url{https://github.com/features/copilot}
\showURL{%
\tempurl}


\bibitem[int(2021)]%
        {intellicode}
Microsoft \bibinfo{year}{Accessed 2021}\natexlab{}.
\newblock \bibinfo{booktitle}{\emph{Microsoft Intellicode}}.
\newblock Microsoft.
\newblock
\urldef\tempurl%
\url{https://visualstudio.microsoft.com/services/intellicode}
\showURL{%
\tempurl}


\bibitem[goo(2021)]%
        {googleblog}
Google \bibinfo{year}{Accessed 2021}\natexlab{}.
\newblock \bibinfo{booktitle}{\emph{ML Enhanced Code Completion}}.
\newblock Google.
\newblock
\urldef\tempurl%
\url{https://ai.googleblog.com/2022/07/ml-enhanced-code-completion-improves.html}
\showURL{%
\tempurl}


\bibitem[cod(2023)]%
        {codewhisperer}
Amazon \bibinfo{year}{Accessed 2023}\natexlab{}.
\newblock \bibinfo{booktitle}{\emph{Amazon CodeWhisperer}}.
\newblock Amazon.
\newblock
\urldef\tempurl%
\url{https://aws.amazon.com/codewhisperer}
\showURL{%
\tempurl}


\bibitem[Cod(2023)]%
        {CodeiumAmazonComparison}
 \bibinfo{year}{Accessed 2023}\natexlab{}.
\newblock \bibinfo{title}{{Codeium vs Amazon CodeWhisperer}}.
\newblock
\newblock
\urldef\tempurl%
\url{https://codeium.com/compare/comparison-codewhisperer-codeium}
\showURL{%
\tempurl}


\bibitem[CoP(2023)]%
        {CoPilotLatency}
 \bibinfo{year}{Accessed 2023}\natexlab{}.
\newblock \bibinfo{title}{{Smarter, more efficient coding: GitHub Copilot goes
  beyond Codex with improved AI model}}.
\newblock
\newblock
\urldef\tempurl%
\url{https://github.blog/2023-07-28-smarter-more-efficient-coding-github-copilot-goes-beyond-codex-with-improved-ai-model/}
\showURL{%
\tempurl}


\bibitem[xfo(2023)]%
        {xformers}
GitHub \bibinfo{year}{Accessed 2023}\natexlab{}.
\newblock \bibinfo{booktitle}{\emph{xformers}}.
\newblock GitHub.
\newblock
\urldef\tempurl%
\url{https://github.com/facebookresearch/xformers/tree/main/xformers}
\showURL{%
\tempurl}


\bibitem[Barke et~al\mbox{.}(2023)]%
        {Barke2023OOPSLA1}
\bibfield{author}{\bibinfo{person}{Shraddha Barke}, \bibinfo{person}{Michael~B.
  James}, {and} \bibinfo{person}{Nadia Polikarpova}.}
  \bibinfo{year}{2023}\natexlab{}.
\newblock \showarticletitle{Grounded Copilot: How Programmers Interact with
  Code-Generating Models}.
\newblock  \bibinfo{volume}{7}, \bibinfo{number}{OOPSLA1}, Article
  \bibinfo{articleno}{78} (\bibinfo{date}{apr} \bibinfo{year}{2023}),
  \bibinfo{numpages}{27}~pages.
\newblock
\urldef\tempurl%
\url{https://doi.org/10.1145/3586030}
\showDOI{\tempurl}


\bibitem[Bird et~al\mbox{.}(2023)]%
        {Bird2023Queue}
\bibfield{author}{\bibinfo{person}{Christian Bird}, \bibinfo{person}{Denae
  Ford}, \bibinfo{person}{Thomas Zimmermann}, \bibinfo{person}{Nicole
  Forsgren}, \bibinfo{person}{Eirini Kalliamvakou}, \bibinfo{person}{Travis
  Lowdermilk}, {and} \bibinfo{person}{Idan Gazit}.}
  \bibinfo{year}{2023}\natexlab{}.
\newblock \showarticletitle{Taking Flight with Copilot: Early insights and
  opportunities of AI-powered pair-programming tools}.
\newblock \bibinfo{journal}{\emph{Queue}} \bibinfo{volume}{20},
  \bibinfo{number}{6} (\bibinfo{date}{jan} \bibinfo{year}{2023}),
  \bibinfo{pages}{35–57}.
\newblock
\showISSN{1542-7730}
\urldef\tempurl%
\url{https://doi.org/10.1145/3582083}
\showDOI{\tempurl}


\bibitem[Bruch et~al\mbox{.}(2009)]%
        {bruch2009learning}
\bibfield{author}{\bibinfo{person}{Marcel Bruch}, \bibinfo{person}{Martin
  Monperrus}, {and} \bibinfo{person}{Mira Mezini}.}
  \bibinfo{year}{2009}\natexlab{}.
\newblock \showarticletitle{Learning from examples to improve code completion
  systems}. In \bibinfo{booktitle}{\emph{Proceedings of the 7th joint meeting
  of the European software engineering conference and the ACM SIGSOFT symposium
  on The foundations of software engineering (ESEC/FSE '09)}}.
\newblock


\bibitem[Dao et~al\mbox{.}(2022)]%
        {dao2022flashattention}
\bibfield{author}{\bibinfo{person}{Tri Dao}, \bibinfo{person}{Daniel~Y. Fu},
  \bibinfo{person}{Stefano Ermon}, \bibinfo{person}{Atri Rudra}, {and}
  \bibinfo{person}{Christopher Ré}.} \bibinfo{year}{2022}\natexlab{}.
\newblock \bibinfo{title}{FlashAttention: Fast and Memory-Efficient Exact
  Attention with IO-Awareness}.
\newblock
\newblock
\showeprint[arxiv]{2205.14135}~[cs.LG]


\bibitem[Fried et~al\mbox{.}(2023)]%
        {fried2023incoder}
\bibfield{author}{\bibinfo{person}{Daniel Fried}, \bibinfo{person}{Armen
  Aghajanyan}, \bibinfo{person}{Jessy Lin}, \bibinfo{person}{Sida Wang},
  \bibinfo{person}{Eric Wallace}, \bibinfo{person}{Freda Shi},
  \bibinfo{person}{Ruiqi Zhong}, \bibinfo{person}{Wen tau Yih},
  \bibinfo{person}{Luke Zettlemoyer}, {and} \bibinfo{person}{Mike Lewis}.}
  \bibinfo{year}{2023}\natexlab{}.
\newblock \bibinfo{title}{InCoder: A Generative Model for Code Infilling and
  Synthesis}.
\newblock
\newblock
\showeprint[arxiv]{2204.05999}~[cs.SE]


\bibitem[Kim et~al\mbox{.}(2021)]%
        {kim21code}
\bibfield{author}{\bibinfo{person}{Seohyun Kim}, \bibinfo{person}{Jinman Zhao},
  \bibinfo{person}{Yuchi Tian}, {and} \bibinfo{person}{Satish Chandra}.}
  \bibinfo{year}{2021}\natexlab{}.
\newblock \showarticletitle{Code Prediction by Feeding Trees to Transformers}.
  In \bibinfo{booktitle}{\emph{2021 IEEE/ACM 43rd International Conference on
  Software Engineering (ICSE)}}. \bibinfo{pages}{150--162}.
\newblock
\urldef\tempurl%
\url{https://doi.org/10.1109/ICSE43902.2021.00026}
\showDOI{\tempurl}


\bibitem[Murali et~al\mbox{.}(2024)]%
        {codecompose2024}
\bibfield{author}{\bibinfo{person}{Vijayaraghavan Murali},
  \bibinfo{person}{Chandra Maddila}, \bibinfo{person}{Imad Ahmad},
  \bibinfo{person}{Michael Bolin}, \bibinfo{person}{Daniel Cheng},
  \bibinfo{person}{Negar Ghorbani}, \bibinfo{person}{Renuka Fernandez},
  \bibinfo{person}{Nachiappan Nagappan}, {and} \bibinfo{person}{Peter~C
  Rigby}.} \bibinfo{year}{2024}\natexlab{}.
\newblock \showarticletitle{AI-assisted Code Authoring at Scale: Fine-tuning,
  deploying, and mixed methods evaluations}. In
  \bibinfo{booktitle}{\emph{Proceedings of the Foundations of Software
  Engineering (FSE '24)}}.
\newblock


\bibitem[Nguyen and Nadi(2022)]%
        {nguyen2022empirical}
\bibfield{author}{\bibinfo{person}{Nhan Nguyen} {and} \bibinfo{person}{Sarah
  Nadi}.} \bibinfo{year}{2022}\natexlab{}.
\newblock \showarticletitle{An empirical evaluation of GitHub copilot's code
  suggestions}. In \bibinfo{booktitle}{\emph{Proceedings of the 19th
  International Conference on Mining Software Repositories (MSR '22)}}.
\newblock


\bibitem[Proksch et~al\mbox{.}(2015)]%
        {proksch2015intelligent}
\bibfield{author}{\bibinfo{person}{Sebastian Proksch},
  \bibinfo{person}{Johannes Lerch}, {and} \bibinfo{person}{Mira Mezini}.}
  \bibinfo{year}{2015}\natexlab{}.
\newblock \showarticletitle{Intelligent Code Completion with Bayesian
  Networks}.
\newblock \bibinfo{journal}{\emph{ACM Transactions on Software Engineering and
  Methodology (TOSEM)}} \bibinfo{volume}{25}, \bibinfo{number}{1}, Article
  \bibinfo{articleno}{3} (\bibinfo{date}{12} \bibinfo{year}{2015}).
\newblock


\bibitem[Robles and Lanza(2008)]%
        {robles2008program}
\bibfield{author}{\bibinfo{person}{R. Robles} {and} \bibinfo{person}{M.
  Lanza}.} \bibinfo{year}{2008}\natexlab{}.
\newblock \showarticletitle{How Program History Can Improve Code Completion}.
  In \bibinfo{booktitle}{\emph{2008 23rd IEEE/ACM International Conference on
  Automated Software Engineering}}. \bibinfo{pages}{317--326}.
\newblock
\urldef\tempurl%
\url{https://doi.org/10.1109/ASE.2008.42}
\showDOI{\tempurl}


\bibitem[Rozière et~al\mbox{.}(2023)]%
        {roziere2023code}
\bibfield{author}{\bibinfo{person}{Baptiste Rozière}, \bibinfo{person}{Jonas
  Gehring}, \bibinfo{person}{Fabian Gloeckle}, \bibinfo{person}{Sten Sootla},
  \bibinfo{person}{Itai Gat}, \bibinfo{person}{Xiaoqing~Ellen Tan},
  \bibinfo{person}{Yossi Adi}, \bibinfo{person}{Jingyu Liu},
  \bibinfo{person}{Tal Remez}, \bibinfo{person}{Jérémy Rapin},
  \bibinfo{person}{Artyom Kozhevnikov}, \bibinfo{person}{Ivan Evtimov},
  \bibinfo{person}{Joanna Bitton}, \bibinfo{person}{Manish Bhatt},
  \bibinfo{person}{Cristian~Canton Ferrer}, \bibinfo{person}{Aaron
  Grattafiori}, \bibinfo{person}{Wenhan Xiong}, \bibinfo{person}{Alexandre
  Défossez}, \bibinfo{person}{Jade Copet}, \bibinfo{person}{Faisal Azhar},
  \bibinfo{person}{Hugo Touvron}, \bibinfo{person}{Louis Martin},
  \bibinfo{person}{Nicolas Usunier}, \bibinfo{person}{Thomas Scialom}, {and}
  \bibinfo{person}{Gabriel Synnaeve}.} \bibinfo{year}{2023}\natexlab{}.
\newblock \bibinfo{title}{Code Llama: Open Foundation Models for Code}.
\newblock
\newblock
\showeprint[arxiv]{2308.12950}~[cs.CL]


\bibitem[Shan et~al\mbox{.}(2022)]%
        {NudgeBot2022FSE}
\bibfield{author}{\bibinfo{person}{Qianhua Shan}, \bibinfo{person}{David
  Sukhdeo}, \bibinfo{person}{Qianying Huang}, \bibinfo{person}{Seth Rogers},
  \bibinfo{person}{Lawrence Chen}, \bibinfo{person}{Elise Paradis},
  \bibinfo{person}{Peter~C. Rigby}, {and} \bibinfo{person}{Nachiappan
  Nagappan}.} \bibinfo{year}{2022}\natexlab{}.
\newblock \showarticletitle{Using nudges to accelerate code reviews at scale}
  \emph{(\bibinfo{series}{ESEC/FSE 2022})}. \bibinfo{publisher}{Association for
  Computing Machinery}, \bibinfo{address}{New York, NY, USA},
  \bibinfo{pages}{472–482}.
\newblock
\showISBNx{9781450394130}
\urldef\tempurl%
\url{https://doi.org/10.1145/3540250.3549104}
\showDOI{\tempurl}


\bibitem[Vaithilingam et~al\mbox{.}(2022)]%
        {vaithilingam2022expectation}
\bibfield{author}{\bibinfo{person}{Priyan Vaithilingam},
  \bibinfo{person}{Tianyi Zhang}, {and} \bibinfo{person}{Elena~L. Glassman}.}
  \bibinfo{year}{2022}\natexlab{}.
\newblock \showarticletitle{Expectation vs. Experience: Evaluating the
  Usability of Code Generation Tools Powered by Large Language Models}. In
  \bibinfo{booktitle}{\emph{Extended Abstracts of the 2022 CHI Conference on
  Human Factors in Computing Systems (CHI EA '22)}}.
\newblock


\bibitem[Zhou et~al\mbox{.}(2022)]%
        {zhou22improving}
\bibfield{author}{\bibinfo{person}{Wen Zhou}, \bibinfo{person}{Seohyun Kim},
  \bibinfo{person}{Vijayaraghavan Murali}, {and} \bibinfo{person}{Gareth~Ari
  Aye}.} \bibinfo{year}{2022}\natexlab{}.
\newblock \showarticletitle{Improving Code Autocompletion with Transfer
  Learning}. In \bibinfo{booktitle}{\emph{2022 IEEE/ACM 44th International
  Conference on Software Engineering: Software Engineering in Practice
  (ICSE-SEIP)}}. \bibinfo{pages}{161--162}.
\newblock
\urldef\tempurl%
\url{https://doi.org/10.1145/3510457.3513061}
\showDOI{\tempurl}


\end{thebibliography}

\end{document}